\documentclass[aps,showpacs,preprintnumbers,superscriptaddress,merge]{revtex4}
\usepackage{hyperref,amsmath,graphicx,subfigure}

\begin{document}
\title{Warming up for Planck} 
\date{\today}
\author{Sam Bartrum}
\email{s.bartrum@sms.ed.ac.uk}

\author{Arjun Berera}
\email{ab@ph.ed.ac.uk}
\affiliation{SUPA, School of Physics and Astronomy, University of Edinburgh, Edinburgh, EH9 3JZ, United Kingdom}

\author{Jo\~{a}o G.~Rosa}
\email{joao.rosa@ua.pt}

\affiliation{Departamento de F\'{\i}sica da Universidade de Aveiro and I3N \\ Campus de Santiago, 3810-183 Aveiro, Portugal}

\preprint{Edinburgh 2013/01}


\begin{abstract}
The recent Planck results and future releases on the horizon present a key opportunity to address a fundamental question in inflationary cosmology of whether primordial density perturbations have a quantum or thermal origin, i.e.~whether particle production may have significant effects during inflation. Warm inflation provides a natural arena to address this issue, with interactions between the scalar inflaton and other degrees of freedom leading to dissipative entropy production and associated thermal fluctuations. In this context, we present relations between CMB observables that can be directly tested against observational data. In particular, we show that the presence of a thermal bath warmer than the Hubble scale during inflation decreases the tensor-to-scalar ratio with respect to the conventional prediction in supercooled inflation, yielding $r< 8|n_t|$, where $n_t$ is the tensor spectral index. Focusing on supersymmetric models at low temperatures, we determine consistency relations between the observables characterizing the spectrum of adiabatic scalar and tensor modes, both for generic potentials and particular canonical examples, and which we compare with the WMAP and Planck results. Finally, we include the possibility of producing the observed baryon asymmetry during inflation through dissipative effects, thereby generating baryon isocurvature modes that can be easily accommodated by the Planck data.

\end{abstract}
\maketitle


\section{Introduction}

The recently announced results of the Planck satellite \cite{Ade:2013rta} seem to paint a simple picture of the early universe, with the observed temperature anisotropies in the Cosmic Microwave Background (CMB) being, at least on small scales, well described by a primordial spectrum of nearly scale invariant, gaussian and adiabatic fluctuations evolving in a $\Lambda$CDM cosmological background. Although the data can be described by a single scalar field slowly rolling down a nearly-flat potential dominating the early stages of an otherwise empty universe, one may wonder whether this is the only simple or even the most likely description of the early universe in the spirit of Occam's razor. In particular, the Planck data strongly disfavors the simplest monomial potentials in chaotic inflation, which have historically been considered as the simplest renormalizable inflationary scenarios, pointing instead towards plateau-like models. In several cases, the latter require a severe parameter fine-tuning unless a special symmetry is present or, alternatively, include modifications of Einstein's gravity, which are not yet fully understood despite the remarkable agreement with the data. Is Planck really validating the inflationary paradigm or, as suggested by some authors \cite{Ijjas:2013vea}, does this create more problems than it solves? One may also wonder whether a completely isolated inflaton field is the most natural scenario, or if the inclusion of (renormalizable) interactions with other fields may also be in agreement with the Planck results.

A fundamental question to address is whether the data really establishes an inflationary universe in a true vacuum state or if more general statistical distributions may yield equally good or even better descriptions of the observational results. A particularly simple and natural possibility is the presence of a thermal state during inflation, which despite the accelerated expansion rate may be sustained if for example the inflaton itself can act as source of particles that quickly thermalize. Although the details of particle production may differ depending on the nature and structure of the interactions between the inflaton and other degrees of freedom, this generic inflationary paradigm is known as warm inflation \cite{Berera:1996nv,Berera:1995ie,Berera:1995wh,Berera:1999ws,BasteroGil:2009ec,Moss:2008yb} (see also \cite{Fang:1980wi, Moss:1985wn, Yokoyama:1987an}; in an observational context see \cite{Hall:2004ab,Taylor:2000ze,Ramos:2013nsa}), with the dynamics of the background field and associated perturbations being governed by fluctuation-dissipation dynamics. 

Warm inflation thus provides a natural background to test alternatives to the standard inflationary paradigm and address the general question of whether primordial density perturbations have a quantum or thermal origin. The announcement of the Planck results and the prospects for new data releases concerning CMB polarization thus motivate a closer look at this issue, in particular determining which observables and relations between them are suitable to address this issue in the light of the new and future observational results. 

Warm inflation can be successfully realized through a generic supersymmetric two-stage model where fields that couple directly to the inflaton acquire a large tree level mass due to the inflaton's large vacuum expectation value (vev), but may subsequently decay into light degrees of freedom, which upon thermalization create a thermal bath concurrent with the inflationary expansion \cite{Berera:2002sp}. If the inflaton vev is sufficiently large, the masses of these mediator fields become heavier than the temperature of the thermal bath, which is known as the low-temperature regime, and thermal corrections to the inflaton potential are thus Boltzmann-suppressed, preserving the tree-level flatness of the potential. Despite supersymmetry being broken due to the finite energy density and temperature during warm inflation, the supersymmetric nature of the model keeps both radiative and thermal corrections to the potential under control. This mechanism leads to dissipative effects that further damp the scalar inflaton's motion, allowing for longer periods of accelerated expansion, and the continuous production of relativistic particles may automatically yield a graceful exit into the standard radiation-dominated cosmology.

As proposed in \cite{Rosa2012}, if the heavy mediators decay through $B$- and $CP$-violating interactions, the production rates of baryons and anti-baryons will be different and a baryon asymmetry may be produced during inflation, a mechanism known as warm baryogenesis. Dissipation is naturally an out-of-equilibrium process, with annihilations of the light degrees of freedom being inefficient in repopulating the classical background condensate, thus satisfying the Sakharov conditions. The structure of the interactions closely resembles that of GUT baryogenesis models, similarly requiring at least two heavy mediators with different masses and complex Yukawa couplings, and in fact the two-stage superpotential is a natural feature of GUT-like gauge theories and associated D-brane constructions \cite{BasteroGil:2011mr}. 

The asymmetry approaches a steady state solution during inflation, due to the slowly varying temperature of the thermal bath and will freeze out once inflation ends and the Hubble parameter drops sufficiently to make the dissipative mechanism inefficient. The produced asymmetry is suppressed in the low-temperature regime, allowing for $\mathcal{O}(1)$ couplings with CP violating phases and mass differences at the few percent level whilst producing an asymmetry within the observed window. This is in contrast to standard thermal GUT scenarios where heavily suppressed couplings are required to generate the observed asymmetry. Furthermore, standard thermal GUT baryogenesis models require a reheating temperature above the GUT threshold in order to generate a baryon asymmetry from an initial population of heavy fields in thermal equilibrium, which may lead to an overproduction of dangerous GUT relics or gravitinos. On the other hand, the temperature during warm inflation must be below the heavy mass threshold, preventing the production of GUT monopoles and potentially solving the gravitino problem for sufficiently strong dissipation \cite{Bartrum:2012tg}.

As the baryon asymmetry produced through dissipation depends on the inflaton vev and the temperature of the thermal bath, the resulting baryon-to-entropy ratio will exhibit fluctuations that are fully correlated or anti-correlated with the adiabatic inflaton fluctuations and induce baryon isocurvature perturbations that are subsequently imprinted on the CMB. Warm baryogenesis thus avoids the problems that plague standard models of thermal GUT baryogenesis whilst simultaneously being a testable model of baryogenesis.

In warm inflation, the observables that characterize the density and tensor perturbations spectrum depend not only on the slow-roll parameters, as in cold inflation, but also on the value of the dissipative coefficient when the relevant scales exit the horizon during inflation, $Q=\Upsilon/3H$, and on the number of relativistic degrees of freedom, $g_*$. In general, this yields too many parameters to allow one to fully relate the observable quantities. On the other hand, the generation of baryon isocurvature perturbations provides an additional observable and, furthermore, canonical potentials used in inflationary model building are characterized by only two or three parameters. As we will show, this will allow us to derive several consistency relations that can be directly tested against observational results.

In this work, we focus on consistency relations between observables based on the above mentioned two-stage supersymmetric warm inflation model, in the low-temperature regime, assuming the observed baryon asymmetry is produced during inflation and that adiabatic perturbations originate solely from thermal fluctuations of the inflaton field. Here, we anticipate the main results of this work, which are the consistency relations for weak and strong dissipation, $Q\ll1$ and $Q\gg1$, respectively, that hold for any generic inflationary potential. These relate the scalar and tensor spectral indices, $n_s$ and $n_t$, the associated running parameters, $n_s'$ and $n_t'$ as well as the running-of-the-running of the scalar index $n_s''$, the tensor-to-scalar ratio, $r$, the fraction of baryon isocurvature perturbations, $B_B$ and its spectral index, $n_{iso}$:
\begin{eqnarray} \label{consistency_general}
B_B&=&4n_t-9(n_s-1)+\frac{8(n_s-1)^2-4n_s'}{n_s-1+n_t} ~, \hspace{7.75cm} Q\ll 1~, \nonumber\\
n_{iso}&=&\frac{1}{4}\Bigg(5(n_s-1)-4n_t+\frac{8(n_s-1)(n_s'-2(n_s-1)^2)}{(n_s-1+n_t)^2}+\frac{6((n_s-1)^2-2n_s')}{n_s-1+n_t} \nonumber \\
&+&\frac{-7+7n_s^3+16n_s''-3n_t-12n_s'(3+n_t)-3n_s^2(7+n_t)+3n_s(7+12n_s'+2n_t)}{1+n_s^2+4n_s'-n_t(5+4n_t)+n_s(5n_t-2)}\Bigg)~, \hspace{1.23cm} Q\ll 1~, \nonumber \\
n_{iso}&=&\frac{n_t^2\left[-32((n_s-1)^2+4n_s')+4(n_s-1)n_t+9n_t^2\right]+2n_t(2(1-n_s)+45n_t)n_t'-3n_t'^2}{8n_t\left[8n_t(1-n_s)+3n_t^2-3n_t'\right]}~,  \qquad Q\gg1~.
\end{eqnarray}
We will place these relations in context below and apply them to some canonical potentials. These relations correspond to the dissipation coefficient obtained in the low-temperature regime through the excitation of heavy virtual modes coupled directly to the inflaton, which is the most well-studied case in the literature (see e.g. \cite{BasteroGil:2009ec, Bartrum:2012tg, Cerezo:2012ub} for recent studies of warm inflation dynamics in this scenario). In this sense, these results cannot be used to probe the full warm inflation paradigm. One should note that, in the low-temperature regime, the excitation of on-shell models may give the dominant contribution to the dissipation coefficient for sufficiently small Yukawa couplings \cite{Rosa2012}, a scenario that is, however, outside the scope of this work.

On the other hand, as we show below, in warm inflation models we find:
\begin{eqnarray} \label{inequality}
r<8|n_t|~,
\end{eqnarray}
independently of the form of the dissipation coefficient, with $r=8|n_t|$ corresponding to cold single-field models. This is also a generic feature of multiple-field inflationary models \cite{Wands:2007bd}, so that the consistency relations in Eq.~(\ref{consistency_general}) may be crucial in distinguishing these from warm inflation models if $r<8|n_t|$ is observed.

We begin by introducing the relevant dynamics of warm inflation in Section \ref{Dynamics}, before going on to look at the relevant observables in Section \ref{Observables}. We then analyze in more detail a few canonical examples, namely monomial, hybrid and hilltop models, in Section \ref{Models}. In Section \ref{Discussion} we discuss our main results and the prospects for the future releases of CMB data by the Planck collaboration.


\section{Warm Inflation Dynamics}\label{Dynamics}

Warm inflation can be realized in supersymmetric theories through a ``two-stage" generic superpotential of the form:
\begin{equation} \label{superpotential}
W=\frac{g}{2}\Phi X^2+\frac{h_i}{2}XY_i^2+f(\Phi)~,
\end{equation}
where the scalar component of $\Phi$ yields the inflaton field and $f(\Phi)$ is a general holomorphic function describing the self-interactions in  this sector. The $X$ components couple directly to the inflaton and become heavy, while the fields in the $Y$ sector remain light and form the radiation bath. The coupling of the inflaton to heavy fields which decay into light degrees of freedom then induces dissipative particle production, yielding an additional source of friction in the inflaton equation of motion and a source of quasi-thermal radiation:
\begin{align}
\ddot{\phi}+(3H+\Upsilon)\dot{\phi}+V_{\phi}=0~, \hspace{1cm}
\dot{\rho}_R+4H\rho_R=3HQ\dot{\phi}^2~,
\end{align}
where $V_{\phi}$ indicates the derivative of the potential with respect to the inflaton field, $\Upsilon$ is the dissipative coefficient and $\rho_R$ the radiation energy density. It has been recently pointed out that bulk viscous effects in the quasi-thermal bath may modify the dynamics and prolong inflation \cite{BasteroGil:2012zr}, although we will not consider this case in our discussion. Accelerated expansion occurs in the slow-roll regime, where the equations of motion reduce to:
\begin{align}
3H(1+Q)\dot{\phi}\approx-V_{\phi}~, \hspace{1cm}
4\rho_R\approx3Q\dot{\phi}^2~,
\end{align}
with $Q=\Upsilon/3H$ as defined above. In warm inflation, slow-roll is achieved if the following conditions are satisfied:
\begin{equation}
\epsilon_{\phi} = \frac{m_p^2}{2}\left(\frac{V_{\phi}}{V}\right)^2 \ll1+Q~, \hspace{0.5cm} |\eta_{\phi}|= m_p^2\left(\frac{|V_{\phi\phi}|}{V}\right)\ll1+Q~,\hspace{0.5cm} \sigma_{\phi}=m_p^2\left(\frac{V_{\phi}}{\phi V}\right)\ll1+Q~,
\end{equation}
where $m_p$ denotes the reduced Planck mass and the last condition ensures a slow variation of the dissipation coefficient given below. Other slow-roll parameters ensure that the radiation is not redshifted, that dissipation does not cause the radiation to dominate too soon and that thermal and quantum corrections to the potential are not too large (see e.g.~\cite{BasteroGil:2009ec}). In the low-temperature regime, the form of the dissipative coefficient depends on whether the mediator fields in the $X$ sector are produced dominantly on- or off-shell \cite{Rosa2012}. We will focus on the latter case, although the production of on-shell modes gives a promising avenue for model building that has yet to be analyzed in detail. The dominant source of dissipation then corresponds to the production and decay of virtual scalar modes \cite{BasteroGil:2010pb, Rosa2012}, giving:
\begin{equation} \label{dissipation_coeff}
\Upsilon \approx C_{\phi}\frac{T^3}{\phi^2}~,
\end{equation}
where the constant $C_{\phi}$ depends on the field multiplicities and couplings in the light and heavy sectors \cite{Rosa2012}. From the slow-roll equations we can relate $Q$ and $\phi$ via:
\begin{equation}
Q^{1/3}(1+Q)^2=2\epsilon_{\phi}\left(\frac{C_{\phi}}{3}\right)^{1/3}\left(\frac{C_{\phi}}{4C_R}\right)\left(\frac{H}{\phi}\right)^{8/3}\left(\frac{m_p}{H}\right)^2~.
\end{equation}
Using these relations, the evolution of the dynamical quantities can be written in terms of the slow-roll parameters:
\begin{align}
\frac{\phi'}{\phi}&=-\frac{\sigma_{\phi}}{(1+Q)}~, \label{phiev}\\
Q'&=\frac{Q}{(1+7Q)}(10\epsilon_{\phi}-6\eta_{\phi}+8\sigma_{\phi})~, \\
\left(\frac{T}{H}\right)'&=\frac{2}{(1+7Q)}\left(\frac{2+4Q}{1+Q}\epsilon_{\phi}-\eta_{\phi}+\frac{1-Q}{1+Q}\sigma_{\phi}\right)\left(\frac{T}{H}\right)~,
\end{align}
where the prime indicates a derivative with respect to the number of e-folds, $dN_e=Hdt$. Note that  Eq.~(\ref{phiev}) is independent of the form of the dissipative coefficient.

For $T>H$, the dominant contribution to the primordial perturbation spectrum are thermal fluctuations of the inflaton field, as opposed to the conventional quantum fluctuations in cold inflation models. Upon exiting the horizon these thermal fluctuations freeze out as classical perturbations and during slow-roll the amplitude of the curvature perturbation power spectrum is given by \cite{Berera:1995wh,Hall:2003zp}:
\begin{equation}\label{Curvature}
P_{\mathcal{R}}^{1/2}\approx\left(\frac{H}{2\pi}\right)\left(\frac{3H^2}{V_{\phi}}\right)(1+Q)^{5/4}\left(\frac{T}{H}\right)^{1/2}~,
\end{equation}
where all quantities are evaluated at horizon-crossing. Note that the cold inflation limit is obtained when $T/H\rightarrow 1$ and $Q\rightarrow 0$, i.e.~in the absence of dissipation the temperature corresponds to that of the cosmological de Sitter horizon. Eq.~(\ref{Curvature}) is strictly valid for a dissipation coefficient that is independent of the temperature of the radiation bath, whereas here we consider the $T$-dependent coefficient obtained in the supersymmetric two-stage model described above. In this case the inflaton and radiation perturbations can become strongly coupled in the regime $Q\gg1$, which leads to an additional growth of the fluctuations before freeze out \cite{Graham:2009bf}. Although this has a negligible effect for weak dissipation, where Eq.~(\ref{Curvature}) holds, this enhances the amplitude of the scalar power spectrum for strong dissipation and may also modify the associated spectral index depending on the variation of the dissipative ratio $Q$ at horizon crossing. However, although we expect the radiation bath to remain close to thermal equilibrium, the dissipative nature of the particle production process may induce non-negligible departures from a perfect fluid description. In particular, shear viscous effects have been shown to damp the growth of the coupled inflaton-radiation perturbations, while preserving the background dynamics, such that for sufficiently large shear viscosity one recovers the form of the scalar power spectrum for a $T$-independent dissipation coefficient \cite{BasteroGil:2011xd}. Since significant shear viscosities are expected for not too large values of the coupling $h$ determining the self-interactions in the light sector $Y$ \cite{BasteroGil:2011xd}, in the remainder of this work we will thus assume that Eq.~(\ref{Curvature}) yields a sufficiently good approximation to the form of the scalar power spectrum.


\section{Observables}\label{Observables}

The dependence of the curvature power spectrum, Eq.~(\ref{Curvature}), on the scale of the perturbations is given by the scalar spectral index \cite{BasteroGil:2009ec}:
\begin{equation}
n_s-1\equiv\frac{d\ln P_R}{d\ln k} = \frac{1}{(1+Q)(1+7Q)}(-(2+9Q)\epsilon_{\phi}-3Q\eta_{\phi}+(2+18Q)\sigma_{\phi})~.
\end{equation}
The running of the scalar spectral index is given by:
\begin{eqnarray}
n_s' \equiv\frac{dn_s}{dN_e} &=& (1-n_s)\frac{Q'}{1+Q}+\frac{1}{1+Q}(-(2-5A_Q)\epsilon_{\phi}'-3A_Q\eta_{\phi}'+(2+4A_Q)\sigma_{\phi}')\nonumber \\
&+&\frac{Q'}{(1+Q)(1+7Q)^2}(5\epsilon_{\phi}-3\eta_{\phi}+4\sigma_{\phi})~,
\end{eqnarray}
where $A_Q=Q/(1+7Q)$ and primes denotes derivatives with respect to the number of e-folds, in particular:
\begin{equation}
\epsilon_{\phi}'=\frac{2\epsilon_{\phi}}{(1+Q)}(2\epsilon_{\phi}-\eta_{\phi})~,\hspace{1cm}\eta_{\phi}'=\frac{\epsilon_{\phi}}{(1+Q)}(2\eta_{\phi}-\xi_{\phi})~,\hspace{1cm}\sigma_{\phi}'=\frac{\sigma_{\phi}}{(1+Q)}(\sigma_{\phi}+2\epsilon_{\phi}-\eta_{\phi})~,
\end{equation}
with $\xi_{\phi}=2m_p^2(V_{\phi\phi\phi}/V_{\phi})$.

Tensor modes are not affected by dissipation due to the weak coupling of the graviton to the thermal bath and so the tensor perturbation spectrum is governed by quantum fluctuations. The tensor-to-scalar ratio is then modified with respect to cold inflation by an additional thermal factor:
\begin{equation}
r = \left(\frac{H}{T}\right)\frac{16\epsilon_{\phi}}{(1+Q)^{5/2}}~.
\end{equation}
We note that as long as the inflaton is solely responsible for the generation of adiabatic perturbations, which are thermally produced, we may use the slow-roll equations to show that:
\begin{equation}
r=\frac{2}{\pi^2 P_{\mathcal{R}}}\left(\frac{H}{m_p}\right)^2~,
\end{equation}
which is analogous to the cold inflaton result. However, one should note that thermal fluctuations are generically larger than quantum fluctuations, yielding the same amplitude for different field values. In particular, in large field models, one may obtain the observed amplitude for $\phi<m_p$ \cite{BasteroGil:2009ec}.

The scale dependence of the tensor power spectrum is given by the tensor spectral index, which due to the decreasing Hubble parameter is always negative:
\begin{equation}
n_t\equiv\frac{d\ln P_h}{d\ln k}=-\frac{2\epsilon_{\phi}}{1+Q} \le0~,
\end{equation}
where $|n_t|\ll 2$ in order to satisfy the slow-roll conditions. In cold inflation with a single slowly rolling scalar field we thus find $r=8|n_t|$, which is a well-known consistency relation. In warm inflation, the relation between the tensor-to-scalar ratio and the tensor spectral index is, on the other hand, given by: 
\begin{equation}\label{Warmtensor}
r = -\left(\frac{H}{T}\right){8n_t\over (1+Q)^{3/2}}  <8|n_t|~,
\end{equation}
for $T>H$, while one recovers the equality for  $T/H\rightarrow 1$ and $Q\rightarrow 0$. Thus, if a future experiment measures $r$ and $n_t$ and finds $r<8|n_t|$, this could be a hint for a thermal origin of the adiabatic perturbations, although this is also the case for generic multiple field models as mentioned earlier and other observables must be used in order to break this degeneracy.

There is also the possibility that the observed baryon asymmetry is produced during warm inflation through dissipative effects as described earlier. In this scenario, the fluctuations of the inflaton are directly transferred to the baryon-to-entropy ratio which yields baryon isocurvature perturbations that freeze out during inflation and are imprinted on the CMB \cite{Bastero-Gil2011}. These isocurvature perturbations, given by $S_B=\delta\eta_s/\eta_s$, are thus fully correlated or anti-correlated with the adiabatic perturbations generated by the inflaton field. It is convenient to express this in terms of the ratio between the baryon isocurvature perturbations and the curvature perturbation $\zeta$, which for the dissipative coefficient in Eq.~(\ref{dissipation_coeff}) is given by  \cite{Rosa2012}:
\begin{equation}
B_B \equiv\frac{S_B}{\zeta}= \frac{2}{(1+Q)^2(1+7Q)}(2\eta_{\phi}(1+Q)-\epsilon_{\phi}(3+Q)-\sigma_{\phi}(3+5Q))~.
\end{equation}
Whether isocurvature perturbations are correlated or anti-correlated with adiabatic perturbations depends upon the sign of $B_B$. 

The spectral index of the isocurvature perturbations is in principle another observable that can be used to test the warm baryogenesis scenario. For the dissipative coefficient in Eq.~(\ref{dissipation_coeff}) the isocurvature spectral index is given by:
\begin{eqnarray}
n_{iso}&=&\frac{d\ln S_B}{d\ln k} = -\frac{7Q}{(1+7Q)^2}(10\epsilon_{\phi}-6\eta_{\phi}+8\sigma_{\phi}) \nonumber \\
&+&\frac{2\epsilon_{\phi}(2\eta_{\phi}-\xi_{\phi})(1+Q)-2\epsilon_{\phi}(2\epsilon_{\phi}-\eta_{\phi})(3+Q)-\sigma_{\phi}(\sigma_{\phi}+2\epsilon_{\phi}-\eta_{\phi})(3+5Q)}{(1+Q)(2\eta_{\phi}(1+Q)-\epsilon_{\phi}(3+Q)-\sigma_{\phi}(3+5Q))} \nonumber \\
&+&\frac{(2\eta_{\phi}-\epsilon_{\phi}-5\sigma_{\phi})(10\epsilon_{\phi}-6\eta_{\phi}+8\sigma_{\phi})Q}{(1+7Q)(2\eta_{\phi}(1+Q)-\epsilon_{\phi}(3+Q)-\sigma_{\phi}(3+5Q))}-\frac{2\epsilon_{\phi}-\eta_{\phi}}{(1+Q)} \nonumber \\
&+&\frac{\epsilon_{\phi}(2-Q)-\eta_{\phi}(2+5Q)+2\sigma_{\phi}(1+Q)}{2(1+7Q)(1+Q)}~.
\end{eqnarray}
 
This implies that the baryon isocurvature spectral index will typically be proportional to a combination of slow-roll parameters for weak dissipation, which is further suppressed by the dissipative ratio $Q$ for strong dissipation (see below). Hence, we expect this to be small in general, although formally it can take any positive or negative value depending on the form of the inflaton potential. For comparison, in the curvaton scenario the spectral index of the (cold dark matter) isocurvature spectrum is the same as for the adiabatic spectrum (see e.g.~\cite{Gordon:2002gv}), whereas isocurvature modes produced from the quantum fluctuations of axion fields correspond to an almost scale invariant spectrum \cite{Beltran:2006sq}. 

The bounds on anti-correlated cold dark matter isocurvature modes from the 9-year WMAP data analysis (WMAP 9), using the combined WMAP+eCMB+BAO+$H_0$ data and taking into account that $\Omega_{\text{CDM}}/\Omega_{B}\approx5$, yield an upper bound $|B_B|\le0.33$ for $B_B<0$ \cite{Hinshaw:2012fq}. Fully correlated modes have not been analyzed with WMAP 9 data, with an earlier analysis yielding $B_B<0.43$ \cite{Gordon:2002gv}.

In its recent data release, Planck has placed new bounds on the presence of a combination of cold dark matter and baryon isocurvature modes (CDI), which are indistinguishable from the CMB point of view. For a general admixture of adiabatic and isocurvature perturbations, independently of their correlation or the latter's spectral index, the Planck results yield $\beta_{\text{iso}}<0.075$ for the CDI fraction at the comoving wavenumber $k_{\text{low}}=0.002\ \mathrm{Mpc}^{-1}$ \cite{Ade:2013rta}. This translates into a bound $|B_B|\lesssim1.5$, which is surprisingly less constraining than the WMAP results. According to the Planck collaboration, this is related to the fact that the data actually prefers models with a significant contribution of CDI (or neutrino density isocurvature modes), driven mainly by the amplitude deficit observed at low multipoles. Although no definite evidence for isocurvature modes was found in the data, it nevertheless allows for a substantially large fraction of isocurvature modes.
 
Given the model-dependence of the sign of $B_B$ and the value of $n_{iso}$ discussed above, in our analysis below we will include the general bound from Planck and the earlier WMAP bounds to illustrate our results. One should nevertheless keep in mind that, for particular potentials, $n_{iso}$ will be related to $n_s-1$ and more stringent bounds should apply as for the axion and curvaton cases. For example, for monomial potentials, $V(\phi)\propto \phi^n$, and a dissipative coefficient of the form Eq.~(\ref{dissipation_coeff}), we find:
\begin{eqnarray}
\frac{n_{iso}}{n_s-1}=\frac{1}{2}-\frac{2}{n-2}~, \hspace{2cm} Q\ll 1~, \nonumber \\
\frac{n_{iso}}{n_s-1}=\frac{42-11n}{84-30n}~, \hspace{2cm} Q\gg1~.
\end{eqnarray}
We would like to emphasize that the warm baryogenesis scenario may generate isocurvature perturbations which are either correlated or anti-correlated with adiabatic perturbations, depending on the inflaton potential, whereas the recent literature has focused mostly on anti-correlated and uncorrelated modes, motivated by curvaton and axion models. We hope that this work motivates the need to look for fully correlated isocurvature modes in future data. 

In general, for the low-temperature dissipative coefficient in Eq.~(\ref{dissipation_coeff}), the expressions for the observables are too complicated to yield simple  relations between them. If, however, we look at the strong ($Q\gg1$) and weak ($Q\ll1$) dissipative regimes separately, these expressions simplify considerably and such consistency relations can be obtained. To illustrate these relations, we will use the 68\% and 95\% C.L. contours in the $(n_s,n_s')$ plane obtained from the Planck data for a $\Lambda$CDM cosmology with a running scalar spectral index with and without tensor perturbations \cite{Ade:2013rta}.

\subsubsection{Weak dissipation}

In the weak dissipative limit the observables reduce to:
\begin{align}
&n_s-1\approx -2\epsilon_{\phi}+2\sigma_{\phi}~, \hspace{1cm}
n_s'\approx -2\epsilon_{\phi}'+2\sigma_{\phi}'~, \hspace{1cm}
n_s''\approx -2\epsilon_{\phi}''+2\sigma_{\phi}''~,\hspace{1cm}
r\approx \frac{16\epsilon_{\phi}}{Q^{1/3}}\left(\frac{16\pi^2P_{\mathcal{R}}C_R}{3}\right)^{1/3}~, \nonumber \\
&n_t\approx -2\epsilon_{\phi}~,\hspace{1cm}
B_B\approx 2(2\eta_{\phi}-3\epsilon_{\phi}-3\sigma_{\phi})~, \hspace{1cm} n_{iso}\approx\frac{9\epsilon_{\phi}^2+2\epsilon_{\phi}(\xi_{\phi}-4\eta_{\phi}+3\sigma_{\phi})+\sigma_{\phi}(6\sigma_{\phi}-5\eta_{\phi})}{3\epsilon_{\phi}-2\eta_{\phi}+3\sigma_{\phi}}~.
\end{align}
From these expressions we can derive two consistency relations for an arbitrary inflationary potential with the low-temperature dissipative coefficient in the weak dissipative regime, $Q\ll1$:
\begin{equation}\label{LowQcons}
B_B=4n_t-9(n_s-1)+\frac{8(n_s-1)^2-4n_s'}{n_s-1+n_t}~,
\end{equation}
\begin{eqnarray}\label{nisolowQ}
n_{iso}&=&\frac{1}{4}\Bigg(5(n_s-1)-4n_t+\frac{8(n_s-1)(n_s'-2(n_s-1)^2)}{(n_s-1+n_t)^2}+\frac{6((n_s-1)^2-2n_s')}{n_s-1+n_t} \nonumber \\
&+&\frac{-7+7n_s^3+16n_s''-3n_t-12n_s'(3+n_t)-3n_s^2(7+n_t)+3n_s(7+12n_s'+2n_t)}{1+n_s^2+4n_s'-n_t(5+4n_t)+n_s(5n_t-2)}\Bigg)~.
\end{eqnarray}

\begin{figure}[htbp]
\centering
\includegraphics[scale=0.27]{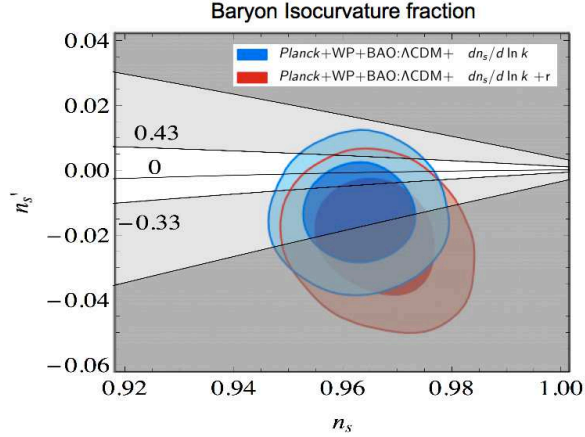}
\caption{Contours of $B_B$ in the ($n_s$,$n_s'$) plane for $Q\ll1$, with $n_t=-0.01$. The dark grey excluded region represents the current bounds on $B_B$ from Planck, while the lighter gray indicates previous limits from WMAP.}
\label{LowQ}
\end{figure}

Figure \ref{LowQ} shows the contours of $B_B$ from Eq.~(\ref{LowQcons}) in the ($n_s$,$n_s'$) plane for $n_t=-0.01$. For $|n_t|\le0.01$ the effect of the tensor index becomes negligible in Eq.~(\ref{LowQcons}) and the parameter space plot tends to this limiting form.  For $n_t\lesssim-0.4$ we do not find any solutions within the $95\%$ C.L., so that this places an upper bound on $|n_t|$ in the weak dissipative regime in the warm baryogenesis scenario.

\subsubsection{Strong dissipation}

In the strong dissipative regime there is no consistency relation involving only $n_s,n_s',r,n_t,B_B$, as there are 5 parameters ($\epsilon_\phi$, $\eta_\phi$, $\sigma_\phi$, $\xi_\phi$, $Q$) for a given value of $g_*$ and only 5 observables. In the strong dissipative limit these are given by:
\begin{align}
&n_s-1\approx\frac{3}{7Q}(6\sigma_{\phi}-3\epsilon_{\phi}-\eta_{\phi})~, \hspace{0.5cm}
n_s'\approx -\frac{3}{49Q^2}  (54\epsilon_{\phi}^2+6\eta_{\phi}^2-2\eta_{\phi}\sigma_{\phi}+6\sigma_{\phi}^2-\epsilon_{\phi}(7\xi_{\phi}+20\eta_{\phi}+48\sigma_{\phi})~, \nonumber \\ 
&r\approx \frac{16\epsilon_{\phi}}{Q^{3}}\left(\frac{16\pi^2P_{\mathcal{R}}C_R}{3}\right)^{1/3}~, \hspace{0.5cm}
n_t\approx-2\frac{\epsilon_{\phi}}{Q}~, \hspace{0.5cm}
B_B\approx \frac{2}{7Q^2}(2\eta_{\phi}-\epsilon_{\phi}-5\sigma_{\phi})~.
\end{align}

However, if we somewhat forward-thinkingly include the running of the tensor spectral index:
\begin{equation}
n_t' = -2\frac{\epsilon'_{\phi}}{1+Q}+2\frac{Q'\epsilon_{\phi}}{(1+Q)^2}= \frac{4\epsilon_{\phi}}{(1+Q)^2(1+7Q)}(-(2+9Q)\epsilon_{\phi}+\eta_{\phi}+4Q(\eta_{\phi}+\sigma_{\phi}))~,
\end{equation}
we obtain the following consistency relation:
\begin{equation}\label{HighQcons}
r=-\frac{1152B_B^2n_t^3}{(n_t(8(1-n_s)+3n_t)-3n_t')^2}\left(\frac{16\pi^2P_{\mathcal{R}}C_R}{3}\right)^{1/3}~.
\end{equation}
We note that the dependence of this relation on the number of relativistic degrees of freedom, $g_*$, is rather mild. For example, for $g_*=\mathcal{O}(10-100)$, the tensor-to-scalar ratio varies only by a factor of $\sim 2$.

In addition, we may consider the isocurvature spectral index, which in the strong dissipative regime is given by:
\begin{equation}
n_{iso}=\frac{27\epsilon_{\phi}^2+28\epsilon_{\phi}\xi_{\phi}-17\epsilon_{\phi}\eta_{\phi}-18\eta_{\phi}^2
-3\epsilon_{\phi}\sigma_{\phi}-29\eta_{\phi}\sigma_{\phi}+80\sigma_{\phi}^2}{14Q(\epsilon_{\phi}-2\eta_{\phi}+5\sigma_{\phi})}~,
\end{equation}
so that we can derive the following consistency relation relating $n_{iso},~n_s,~n_s',n_t$ and $n_t'$:
\begin{equation}\label{nisohighQ}
n_{iso}=\frac{n_t^2\left[-32((n_s-1)^2+4n_s')+4(n_s-1)n_t+9n_t^2\right]+2n_t(2(1-n_s)+45n_t)n_t'-3n_t'^2}{8n_t\left[8n_t(1-n_s)+3n_t^2-3n_t'\right]}~.
\end{equation}
However, since we do not expect the running of the tensor spectral index nor the baryon isocurvature spectral index to be accurately measured in the near future, we will not include the consistency relations in Eq.~(\ref{nisolowQ}), Eq.~(\ref{HighQcons}) and Eq.~(\ref{nisohighQ}) in the remainder of our analysis.


\section{Inflationary models}\label{Models}

In the previous section we have derived general consistency relations between observables that are independent of the form of the inflationary potential in both the weak and the strong dissipative regimes. If we consider specific potentials, the slow-roll parameters are typically no longer independent and we may find stronger consistency relations. In particular, we obtain, in some cases, consistency relations that are independent of the baryon isocurvature perturbations, so that these relations hold even if the baryon asymmetry is not produced during warm inflation. In this section, we thus consider a few canonical models, in particular monomial, hybrid and hilltop potentials.

\subsection{Monomial potentials}\label{Monomials}

Monomial potentials have the generic form $V=V_0(\phi/m_p)^n$. In cold inflation, one typically requires super-planckian field values in order to achieve $\sim40-60$ e-folds of accelerated expansion. For warm inflation, on the other hand, this can be achieved in the sub-planckian regime due to the additional friction. 

The slow-roll parameters for the monomial models are given by:
\begin{equation}
\epsilon_{\phi}=\frac{n^2}{2}\left(\frac{m_p}{\phi}\right)^{2},
\hspace{0.8cm}\eta_{\phi}=n(n-1)\left(\frac{m_p}{\phi}\right)^{2},\hspace{0.8cm}\sigma_{\phi}=\frac{\eta_{\phi}}{(n-1)}~,
\hspace{0.8cm}\xi_{\phi} = 2(n-1)(n-2)\left(\frac{m_p}{\phi}\right)^2~.
\end{equation}

This yields a blue spectral index, $n_s>1$, for the quadratic potential (in contrast to cold inflation), which is disfavoured by the latest Planck data. On the other hand, for the quartic potential one finds a spectral index in excellent agreement with the Planck measurement, even in the weak dissipative regime \cite{BasteroGil:2009ec}.

\subsubsection{Weak dissipative regime}

In the weak dissipative regime, we derive the following consistency relations:
\begin{equation}
B_B=(1-n_s)-\frac{4n_s'}{(n_s-1)}~, \hspace{1cm} n_t = (n_s-1)-\frac{n_s'}{(n_s-1)}~.
\end{equation}

Note that the tensor spectral index is always negative, so that these consistency relations only hold in the regions of the  ($n_s$,$n_s'$) plane where $n_t<0$, i.e. the model can only yield values of the scalar spectral index and running in this region. Also, $n_t\gtrsim -2$ in order to satisfy the slow-roll conditions when the observable scales leave the horizon during inflation.

\begin{figure}[http]
\centering
\subfigure{
\includegraphics[scale=0.27]{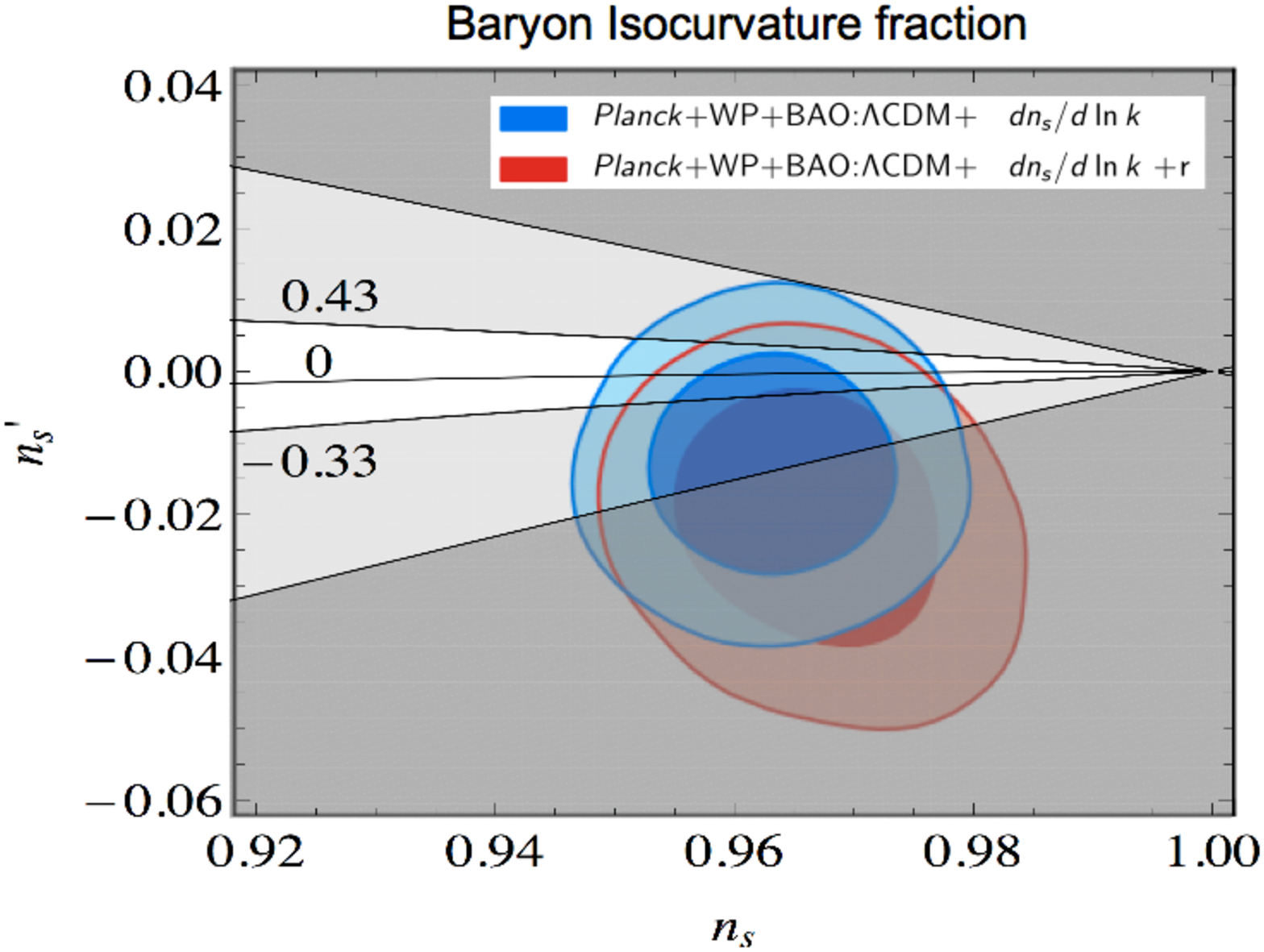} 
}
\subfigure{
\includegraphics[scale=0.27]{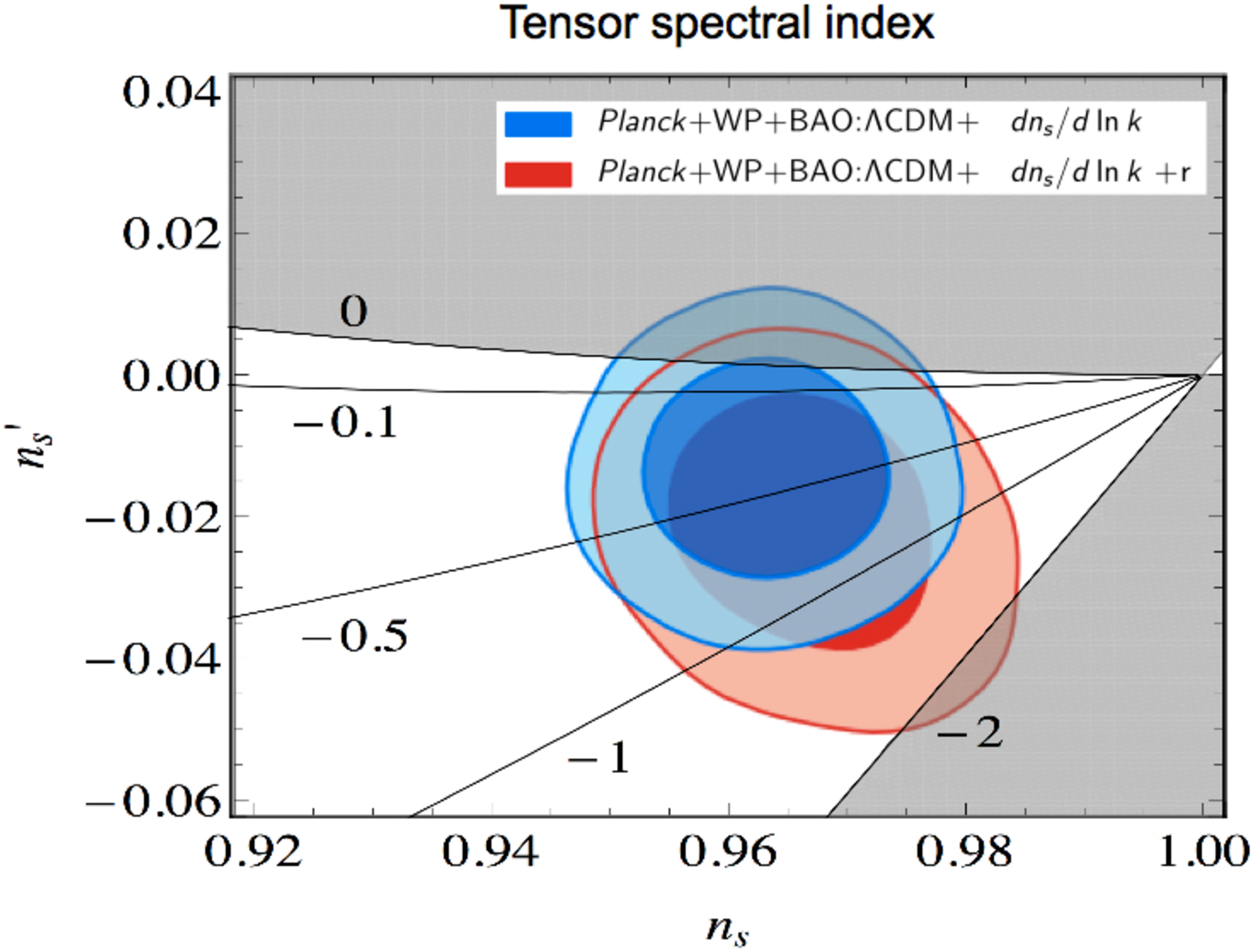}
}
\caption{Contours of $B_B$ (left) and $n_t$ (right) in the ($n_s$,$n_s'$) plane for monomial potentials in the weak dissipative regime.  We exclude the grey (light grey) regions where $B_B$ is outside the observational  bounds for Planck (WMAP) (left) and where the consistency relation yields $n_t>0$ or  $n_t\le-2$ (right).}
\label{MonlowQ}
\end{figure}

As we illustrate in Figure \ref{MonlowQ}, for $n_s\le1$ the allowed regions where $B_B$ is within the observational bounds and  $-2<n_t<0$ overlap with the $95\%$ C.L. region, with $n_t$ falling within the $68\%$ C.L.~contour for $n_t\sim-0.1$.  Also, a positive running would indicate a positive correlation between the isocurvature and adiabatic perturbations with $B_B\gtrsim0.1$ and a tensor spectral index $n_t\gtrsim-0.1$. A negative running may, on the other hand, give anti-correlated isocurvature modes.

\subsubsection{Strong dissipative regime}
 
In the strong dissipative regime the value of $Q$ at horizon-crossing becomes an additional parameter determining the relevant observables compared to the weak dissipative regime. Thus, we may only obtain a single consistency relation:
\begin{equation}\label{MonhighQcons}
n_t = -7\frac{n_s'}{n_s-1}~.
\end{equation}

\begin{figure}[htbp]
\centering
\includegraphics[scale=0.27]{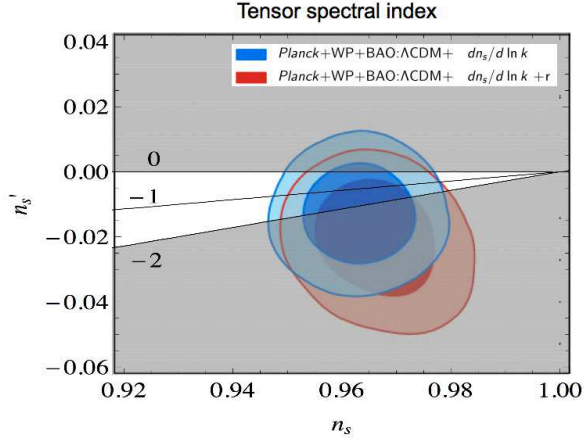}
\caption{Contous of $n_t$ in the ($n_s$,$n_s'$) plane for monomial potentials in the strong dissipative regime. We exclude the grey region where the consistency relation yields $n_t>0$ or  $n_t\le-2$.}
\label{MonhighQ}
\end{figure}

In the strong dissipative regime, $n_t$ is consistent with the $95\%$ C.L. limit and only for larger values, $n_t\sim-0.5$, does $n_t$ fall within the $68\%$ C.L. contour, as shown in Figure \ref{MonhighQ}. For $n_s\le1$ ($\ge1$) the running is necessarily negative (positive), which gives a strong constraint independent of whether the observed baryon asymmetry is produced during warm inflation. In this regime, there is no consistency relation involving $B_B$, but since Eq.~(\ref{MonhighQcons}) only holds for $Q\gg1$ this effectively limits the value of $B_B$ at each point in the ($n_s$,$n_s'$) plane.


\subsection{Hybrid potentials}\label{Hybrid}

We will now consider an example of small field models, hybrid models, where the inflaton couples to one or more `waterfall fields' with a Higgs-like potential. These  fields couple to the inflaton in an analogous fashion to the mediator fields in the supersymmetric model given by Eq.~(\ref{superpotential}), so this provides a natural framework for warm inflation. During inflation, these fields are heavy and overdamped at the origin, yielding a constant cosmological constant that drives the accelerated expansion. This flat potential may be lifted by radiative corrections, soft SUSY breaking mass terms, non-renormalisable interactions or supergravity corrections, typically yielding a leading correction of the form:
\begin{equation}\label{hybridmon}
V=V_0\left(1+\frac{\gamma}{n}\left(\frac{\phi}{m_p}\right)^n\right)~,
\end{equation}
for $n>0$, while for $n=0$ the dominant contribution comes from logarithmic radiative terms:
\begin{equation}\label{hybridlog}
V=V_0\left(1+\gamma\ln\left(\frac{\phi}{m_p}\right)\right)~.
\end{equation}
Here, we will focus on the most common scenarios where either the $n=0$ or the $n=2$ term dominates, both in the strong and weak dissipative regimes. The slow-roll parameters for the hybrid models are given by:
\begin{equation}
\epsilon_{\phi}=\frac{\gamma^2}{2}\left(\frac{m_p}{\phi}\right)^{2-2n}~,
\hspace{0.8cm}\eta_{\phi}=(n-1)\gamma\left(\frac{m_p}{\phi}\right)^{2-n}~,\hspace{0.8cm}\sigma_{\phi}=\frac{\eta_{\phi}}{(n-1)}~,
\hspace{0.8cm}\xi_{\phi} = 2(n-1)(n-2)\left(\frac{m_p}{\phi}\right)^2~,
\end{equation}
where we assumed that, at horizon-crossing, the potential is dominated by the constant term, $V_0$. This limits the values of $\gamma$ and $\phi$ and hence the validity of consistency relations between observables. Note that when the corrections dominate, the potential behaves effectively as a monomial and the relations derived earlier apply. 

For the $n=2$ hybrid model, $n_s>1$ as for the quadratic monomial potential, in both the weak and the strong dissipation regimes at horizon crossing. Since this is disfavoured by the Planck data, we will not consider the consistency conditions in this case, focusing only on the logarithmic hybrid potential.

\subsubsection{$n=0$; Weak dissipative regime}

For the logarithmic hybrid model in Eq.(\ref{hybridlog}), we obtain the following consistency relations: 
\begin{equation}
B_B = 13(1-n_s)\pm8\sqrt{2(n_s-1)^2-n_s'}~,\hspace{1cm} n_t=(1-n_s)\pm\sqrt{2(n_s-1)^2-n_s'}~.
\end{equation}
Note that the `+'  solutions correspond to $\gamma<1$, while the `$-$' solutions hold for $\gamma>1$. We find that only the `$-$' solution has a region of parameter space in which the corrections are small and the consistency relation predictions are compatible with the Planck results.

\begin{figure}[htbp]
\centering
\includegraphics[scale=0.27]{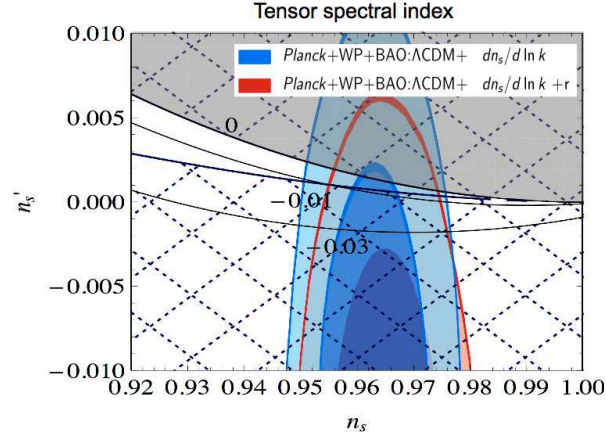} 
\caption{Contours of $n_t$ in the ($n_s$,$n_s'$) plane for the $n=0$ hybrid model in the weak dissipative regime, using the `$-$' solution. We exclude the grey regions where the consistency relation predicts a positive or imaginary $n_t$. In the blue cross-hatched region the radiative corrections dominate, $\gamma|\ln(\phi/m_p)|\ge1$. }
\label{HlowQn0}
\end{figure}

\begin{figure}[htbp]
\centering
\includegraphics[scale=0.27]{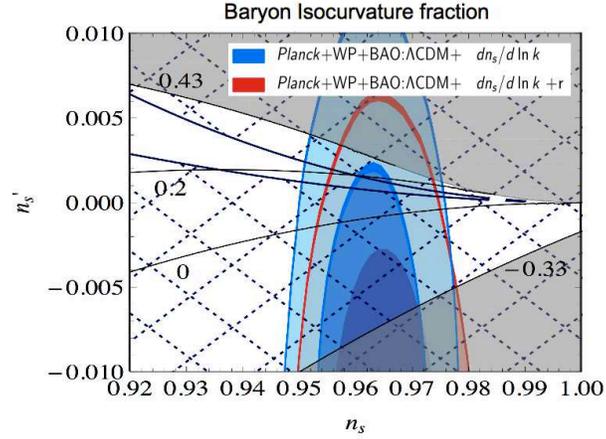} 
\caption{Contours of $B_B$ in the ($n_s$,$n_s'$) plane for the $n=0$ hybrid model in the weak dissipative regime, using the `$-$' solution. The light grey region indicates where $B_B$ exceeds WMAP bounds. In the blue cross-hatched region the radiative corrections dominate, $\gamma|\ln(\phi/m_p)|\ge1$.}
\label{HlowQn02}
\end{figure}

As illustrated in Figures \ref{HlowQn0} and \ref{HlowQn02}, the logarithmic corrections dominate in a large fraction of the parameter space, where the potential behaves effectively as a single monomial. The constant part only dominates in a narrow region of parameter space, where $n_s'\sim 0.005$, $n_t\sim -0.01$ and we find relatively large, positive values of $B_B\sim0.2 - 0.4$. 


\subsubsection{$n=0$; Strong dissipative regime}

\begin{figure}[htbp]
\centering
\includegraphics[scale=0.27]{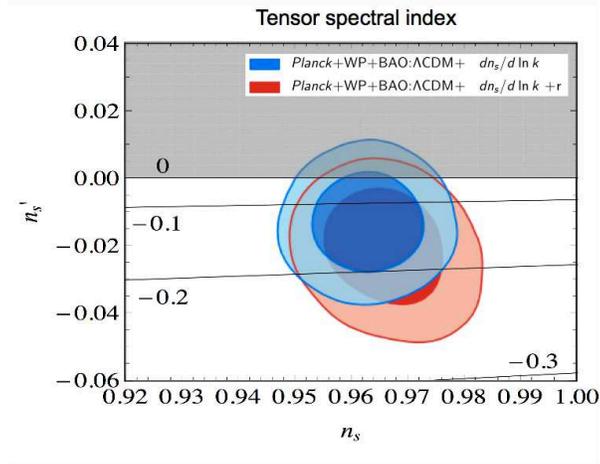}
\caption{Contours of $n_t$ in the ($n_s$,$n_s'$) plane for the hybrid model with $n=0$ in the strong dissipative regime. We exclude the grey region where the consistency relation yields $n_t>0$ or imaginary. }
\label{H0highQ}
\end{figure}

For the logarithmic hybrid model, $n=0$, in the strong dissipative regime we find the following relation:
\begin{equation}\label{nohighQ}
n_t=\frac{1}{9}\left(2(1-n_s)\pm\sqrt{4(n_s-1)^2-126n_s'}\right)~.
\end{equation}
Note that the `$+$' and `$-$' solutions correspond to $\gamma<7/12$ and $\gamma>7/12$, respectively. However, the `$+$' solution only yields $n_t<0$ for $n_s\gtrsim2$ and hence is not consistent within the observationally allowed region. In Figure \ref{H0highQ} we illustrate the contours of $n_t$ for the `$-$' solution.

For this relation, we find that $Q\gg1$ is satisfied and the logarithmic corrections are subdominant within the region where $n_t<0$ for $n_s\ge1$.  For $n_s\le 1$ keeping the logarithmic corrections subdominant gives an upper bound on $n_s'$, for example for the central value of $n_s\sim0.96$ we find $|n_s'|\ge0.005$ with $n_s'\le0$.


\subsection{Hilltop potentials}\label{Hilltop}

Hilltop scenarios are small field models where inflation occurs near a maximum of the potential, being parametrized by:
\begin{equation}
V=V_0\left(1-\frac{M}{2}\left(\frac{\phi}{m_p}\right)^2\right)~.
\end{equation}
Hence, this is equivalent to the $n=2$ hybrid model with a negative mass squared, $\gamma\rightarrow-M$. The consistency relations are thus the same for the hilltop model as for the $n=2$ hybrid model, although the requirement $M(\phi/m_p)^2/2\le1$ or $\ge0$ yields distinct constraints on the parameters.


\subsubsection{Weak dissipative regime}

In the weak dissipative regime we derive the following consistency relation for the hilltop potential:
\begin{equation}
B_B=(1-n_s)-\frac{2n_s'}{n_s-1},\hspace{1cm} n_t=\frac{n_s'}{2(1-n_s)}.
\end{equation}

\begin{figure}[htbp]
\centering
\subfigure{
\includegraphics[scale=0.27]{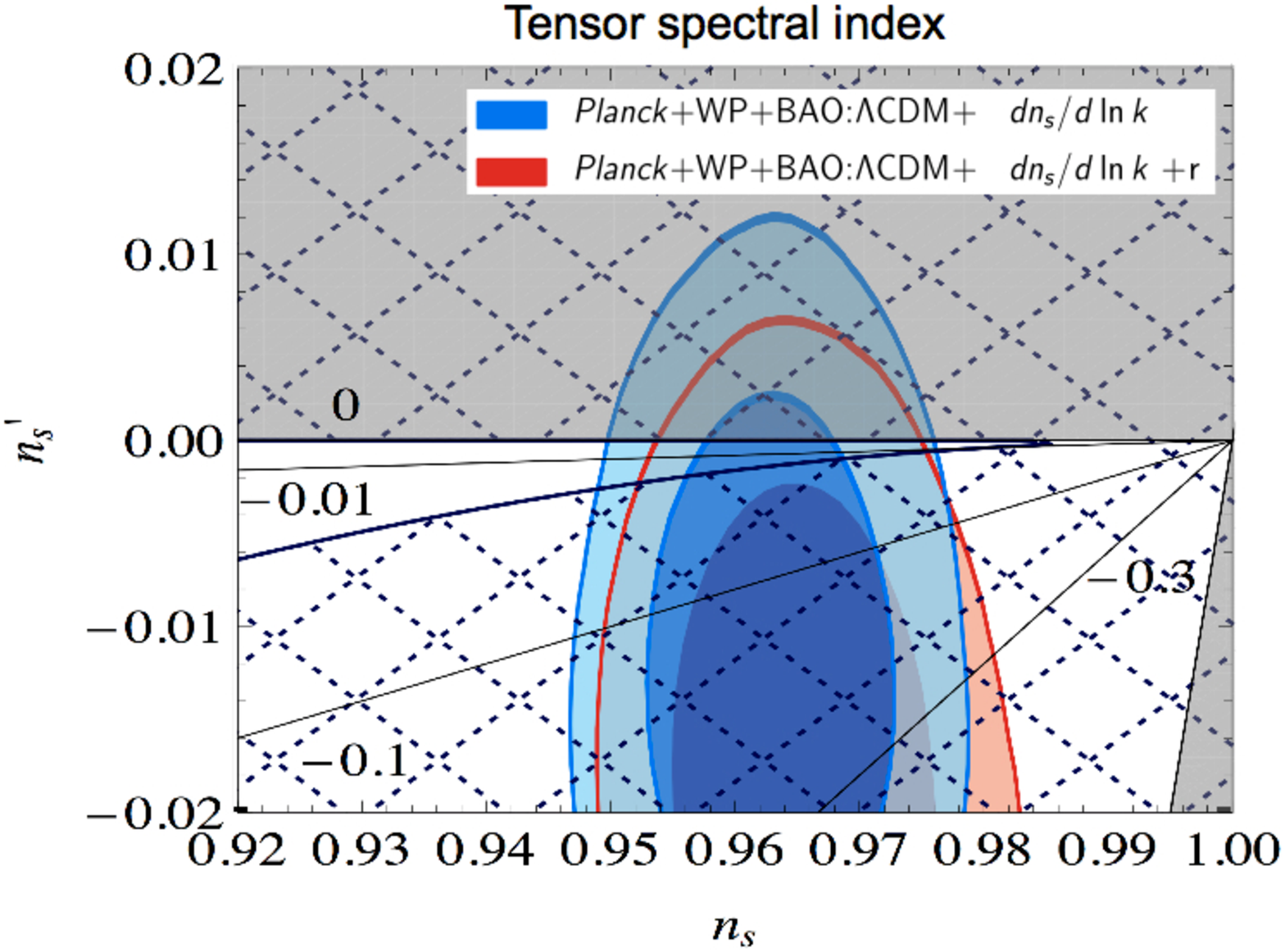}
}
\subfigure{
\includegraphics[scale=0.27]{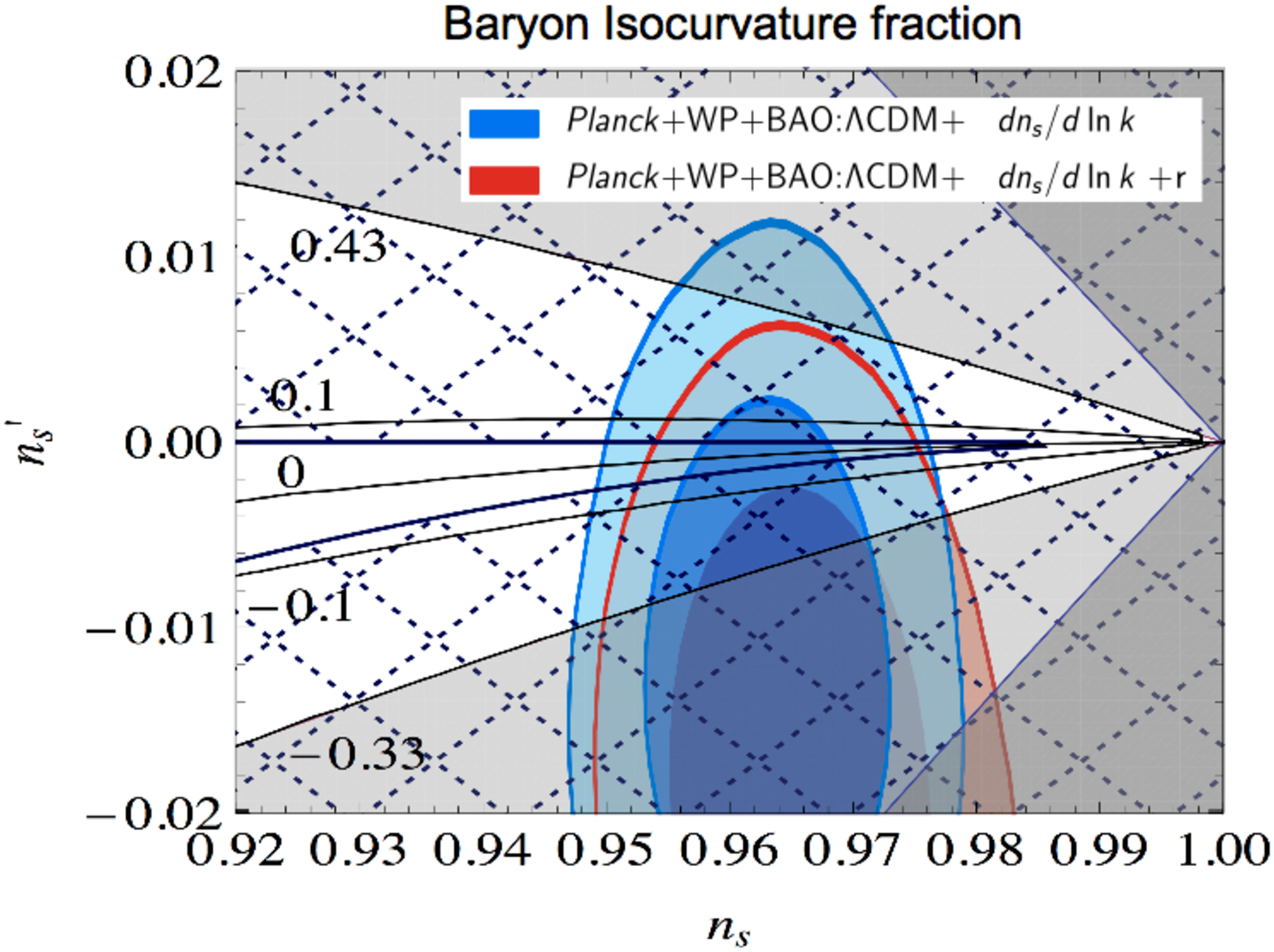}
}
\caption{Contours of $n_t$ (left) and $B_B$ (right)  in the ($n_s$,$n_s'$) plane for the hilltop model in the weak dissipative regime. The consistency relations and exclusions are the same as for the $n=2$ hybrid model, although the region in which the calculation is valid is distinct.}
\label{HilltoplowQ}
\end{figure}

The only simultaneously allowed regions of parameter space for $B_B$ and $n_t$ correspond to $n_s\le1$ with negative running, $n_s'\sim-0.001$,  as shown in Figure \ref{HilltoplowQ}. Within this region we find $n_t\sim-0.01$ and $\sim -0.05 \lesssim B_B \lesssim 0.1$.

\subsubsection{Strong dissipative regime}

In the strong dissipative regime we derive the following consistency relation:
\begin{equation}
n_t=\frac{1}{9}\left(14(1-n_s)\pm\sqrt{154(n_s-1)^2-315n_s'}\right)~.
\end{equation}
However, only the `$-$' solution yields regions of parameter space where the constant term dominates and the consistency relation is compatible with the observational results.

\begin{figure}[htbp]
\centering
\includegraphics[scale=0.27]{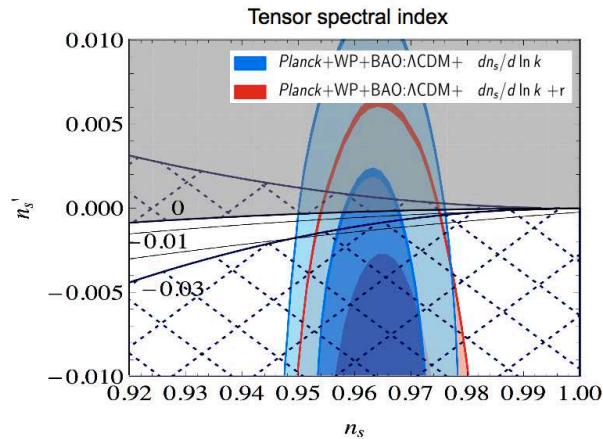}
\caption{Contours of $n_t$ in the ($n_s$,$n_s'$) plane for the hilltop model in the strong dissipative regime. We exclude the grey region where the consistency relation yields $n_t>0$. The blue cross-hatched region indicates where $M(\phi/m_p)^2/2\ge1$ or $<0$.}
\label{HilltophighQ}
\end{figure}

Although the region where the consistency relation is valid is rather small in this case,  it is in an observationally relevant range. In particular, it predicts $n_t\sim-0.02$ with a small but non-vanishing spectral running, $n_s'\sim-0.002$.


\section{Discussion}\label{Discussion}

In this work, we have derived consistency relations between observables for warm inflation models based on supersymmetric two-stage interactions in the low-temperature regime. We have focused on the most well-studied case where dissipation is dominantly generated by the excitation and decay of heavy virtual modes coupled directly to the inflaton. We have obtained relations amongst the parameters characterizing the spectrum of adiabatic perturbations in the scalar and tensor sectors, as well as the fraction of baryon isocurvature perturbations and its spectral index associated to the generation of the observed baryon asymmetry during warm inflation through the dissipative interactions.

The relations between the different observables are, in general, rather intricate and it is not always possible to express them in a simple analytical form. However, we have shown that in the weak and strong dissipative regimes one may obtain generic relations that are independent of the inflationary potential and which constitute the main result of this work, in  Eqs.~(\ref{LowQcons}), (\ref{nisolowQ}), (\ref{HighQcons}) and (\ref{nisohighQ}). 

In addition we have noted that the thermal origin of the adiabatic perturbations modifies the relation between the tensor-to-scalar ratio and the tensor spectral index, giving $r<8|n_t|$, which is always smaller than the corresponding cold inflation prediction. Although this is also a feature of supercooled models with multiple scalar fields, this inequality may be used along with the consistency relations derived in this work to probe the thermal nature of the inflationary state of the Universe.

Prior to this year's release, the bounds on the baryon isocurvature perturbations were expected to increase by an order of magnitude and the tensor-to-scalar ratio was expected to be probed down to $\sim 0.05$ \cite{Planck:2006aa}. Interestingly the bounds on isocurvature appear more relaxed compared to WMAP data, but  as pointed out in \cite{Ade:2013rta} the relaxed constraints arise from the observed deficit in the Sachs-Wolfe plateau and not due to the expected phase shift in the acoustic oscillations which would be a smoking-gun for an isocurvature component. This effect could have several different explanations and is thus not a definite sign of isocurvature modes, although the data allows for a significant admixture of adiabatic and isocurvature components. 

In the warm baryogenesis scenario, the correlation between the produced baryon isocurvature perturbations and the main adiabatic fluctuations, as well as the isocurvature spectral index, depend on the form of the inflationary potential, yielding in general less stringent bounds on the isocurvature fraction than for example in the axion or curvaton scenarios. However, when considering specific inflationary potentials the model predicts a spectral index that can be used to derive stronger bounds on the isocurvature contribution. The Planck collaboration will complete and release its CMB polarization data analysis next year, with a revised analysis of inflationary parameters. One then expects the constraints on the relevant observables to increase significantly and hopefully for new observational signatures to arise as well.  However, even if an isocurvature component is unambiguously found, Planck will be unable to distinguish between cold dark matter and baryon isocurvature perturbations, with a promising observational avenue being future 21 cm fluctuation surveys \cite{Kawasaki:2011ze}.

The Planck results for the scalar spectral index and associated running are even more constraining. With only the WMAP 9-year data the central value of the spectral index is red (blue) without (with) running, although consistent with a scale invariant power spectrum in both cases at the 95$\%$ C.L. A combination of the WMAP data with eCMB, BAO and $H_0$ helps alleviating this degeneracy, at the same time ruling out the Harrison-Zeldovich spectrum at the $95\%$ C.L. and yielding a negative running \cite{Hinshaw:2012fq}. On the other hand, the Planck collaboration has found robust evidence for a red spectral index, with and without including running, thus alleviating this degeneracy. This disfavours the quadratic monomial and hybrid potentials in warm inflation, given that $n_s>1$ in both cases. We note that although in cold inflation the quartic potential is now ruled out, in warm inflation, on the other hand, it shows an impressive agreement with data \cite{BasteroGil:2009ec}.

Warm inflation models may also lead to observable nongaussian signatures in the primordial spectrum from the coupling between the inflaton and radiation fluctuations. These depend on the form and strength of the dissipation coefficient, as well as on the viscous properties of the radiation bath, although in general we may expect $f_{NL}\lesssim 10$ \cite{Gupta:2002kn, Chen:2007gd, Moss:2007cv, *Moss:2011qc}. Planck has so far found no evidence for primordial non-gaussianity, and a dedicated analysis of warm inflationary models for a $T$-independent dissipation coefficient places an upper bound on the dissipative ratio at horizon crossing, $Q\lesssim 400$ \cite{Ade:2013rta}, which represents an improvement of two orders of magnitude with respect to the earlier WMAP bounds. Although this is not directly applicable to the case considered in this work, we expect that further exploration of nongaussianity in general regimes of warm inflation combined with the results presented in this work may help constraining different models of warm inflation in the future.

Many of the consistency relations derived in this work require a knowledge of the properties of the tensor power spectrum, in particular its (running) scale dependence. In this sense, some of our results may only be tested with the next generation of CMB polarization experiments. Nevertheless, we expect Planck's analysis of the CMB polarization spectrum to yield much tighter constraints on cold and warm inflation models next year.

Although our analysis has focused on a particular realization of warm inflation that has been well established in the context of quantum field theory, our results should be put in a wider perspective of an alternative paradigm where thermal or even non-thermal statistical states describe the evolution of the inflationary universe. In particular, we expect other related models to yield relations between CMB observables that can be directly tested against observational data, such as those presented in this work. Hence, we hope that our work motivates further research in this direction from both the observational and theoretical perspectives.


\begin{acknowledgments}
S.B.~and A.B.~are supported by the Science and Technology Facilities Council (United Kingdom). J.G.R.~is supported by the FCT grant SFRH/BPD/85969/2012 (Portugal), as well as partially supported by the grant PTDC/FIS/116625/2010 (Portugal) and the Marie Curie action NRHEP-295189-FP7-PEOPLE-2011-IRSES. J.G.R.~would like to acknowledge the hospitality of the Higgs Centre for Theoretical Physics of the University of Edinburgh during the completion of this work.
\end{acknowledgments}


\bibliographystyle{apsrev} 
                 
\bibliography{ObservablePaper-2.bib}

\end{document}